\documentclass[twocolumn,showpacs,preprintnumbers,amsmath,amssymb,floatfix]{revtex4}

\usepackage{epsfig,psfrag}
\usepackage{dcolumn}
\usepackage{bm}
\usepackage{graphicx}
\usepackage{color}

\begin{document}

\title{Helical Luttinger liquid in topological insulator nanowires }
\author{R. Egger,$^{1,2}$ A. Zazunov,$^1$  and A. Levy Yeyati$^2$}
\affiliation{$^1$Institut f\"ur Theoretische Physik, 
Heinrich-Heine-Universit\"at, D-40225  D\"usseldorf, Germany\\
$^2$Departamento de F{\'i}sica Te{\'o}rica de la Materia Condensada C-V,
Universidad Aut{\'o}noma de Madrid, E-28049 Madrid, Spain }
\date{\today}

\begin{abstract}
We derive and analyze the effective low-energy theory for 
interacting electrons in a cylindrical nanowire made of a 
strong topological insulator. 
Three different approaches provide a consistent
picture for the band structure, where surface states 
forming inside the bulk gap correspond to one-dimensional 
bands indexed by total angular momentum.
When a half-integer magnetic flux pierces the nanowire, we find 
a strongly correlated helical Luttinger liquid 
topologically protected against weak disorder. 
We describe how transport experiments can detect this state.
\end{abstract}

\maketitle

The rich and fascinating physics found in certain spin-orbit coupled
materials exhibiting the ``strong topological insulator'' (TI) phase
currently attracts an enormous amount of attention \cite{hasan}.
In a TI the bulk has a finite gap $\Delta_b$ but topologically 
protected surface modes exist inside the gap.
Using Bi$_2$Se$_3$, which presently is the reference material 
due to its rather large gap, $\Delta_b\approx 0.3$~eV,
surface probe experiments (ARPES, STM) have provided 
clear evidence for the theoretically predicted massless Dirac 
fermion surface state with spin and momentum locked together
\cite{surface}.  However, probing the surface state in
transport experiments still poses a major challenge because  
residual bulk charge carriers -- either related to disorder or due to 
unintentional intrinsic doping -- tend to mask the surface 
contribution even in the cleanest samples so far available 
\cite{butch}.  The surface contribution is easier
to extract experimentally in thin-film geometries \cite{chch} 
or in TI nanowires \cite{exp1,exp2}, where the surface-to-volume
ratio is more advantageous.  In the latter case,
introduction of a magnetic flux $\Phi$ piercing the nanowire 
has allowed to successfully identify the Aharonov-Bohm effect 
caused by the surface state.  

These recent developments clearly demonstrate the need for 
a comprehensive effective low-energy theory of the
electronic properties of TI nanowires, which
we formulate here. Very recent work \cite{ashvin,piet} has
addressed the effect of strong disorder  for the 
noninteracting problem. 
We instead consider the weak disorder limit but take into 
account electron-electron (e-e) interactions 
in a nonperturbative way. 
We obtain the band structure of a cylindrical TI nanowire from
three different approaches:  (i) using the
low-energy approach of Zhang \textit{et al.} \cite{zhang}, (ii)
from the distorted diamond lattice model with spin-orbit
couplings introduced by Fu \textit{et al.} \cite{fu}, and (iii) 
using a surface Dirac fermion theory \cite{ran}. 
Taken together, these calculations draw a consistent picture 
for the surface states inside the bulk gap, even for very thin 
nanowires: a one-dimensional (1D) electron waveguide with 
modes indexed by the half-integer total angular momentum $j$ is formed,
where each mode contains a right- and a left-mover. 
The spin direction is always tangential to the surface and 
perpendicular to the momentum.
For integer flux $\Phi$ (in units of the flux quantum), 
we have an even number of \textit{massive} 1D Dirac fermion modes, 
unlike the case of carbon nanotubes \cite{cnt1,cnt2}.
This allows for impurity backscattering,
and with e-e interactions one has standard disordered Luttinger 
liquid (LL) behavior \cite{gogolin,mora}, where the 
$SU(2)$ spin symmetry is broken.   
The case of half-integer $\Phi$ is more intriguing. Here an emergent
time reversal symmetry (TRS) for the surface states 
results in an odd number of modes topologically protected 
against weak disorder.  With interactions this yields a 
\textit{helical Luttinger liquid}. In the simplest single-mode
case, the spin polarization of a right (left) mover has a counter-clockwise 
(clockwise) orientation around the waist of the
cylinder.  The helical LL has been described previously 
\cite{helical1,helical2}  
as edge mode of the 2D ``quantum spin Hall'' (QSH) topological insulator
realized in HgTe$/$CdTe quantum well structures \cite{konig}.
However, it has been difficult to reveal the QSH helical LL experimentally,
since usually the edges living on 
opposite boundaries both contribute. 
While more complicated setups involving junctions
of different QSH systems have been suggested \cite{helical3},
the situation is unique for a TI nanowire at half-integer
$\Phi$: the fermion doubling theorem
\cite{helical1} is circumvented and, effectively, just one 
QSH edge can be realized.  This simpler realization of a helical LL
should allow for clear signatures in transport experiments. 

\begin{figure}
\scalebox{0.32}{\includegraphics[angle=0]{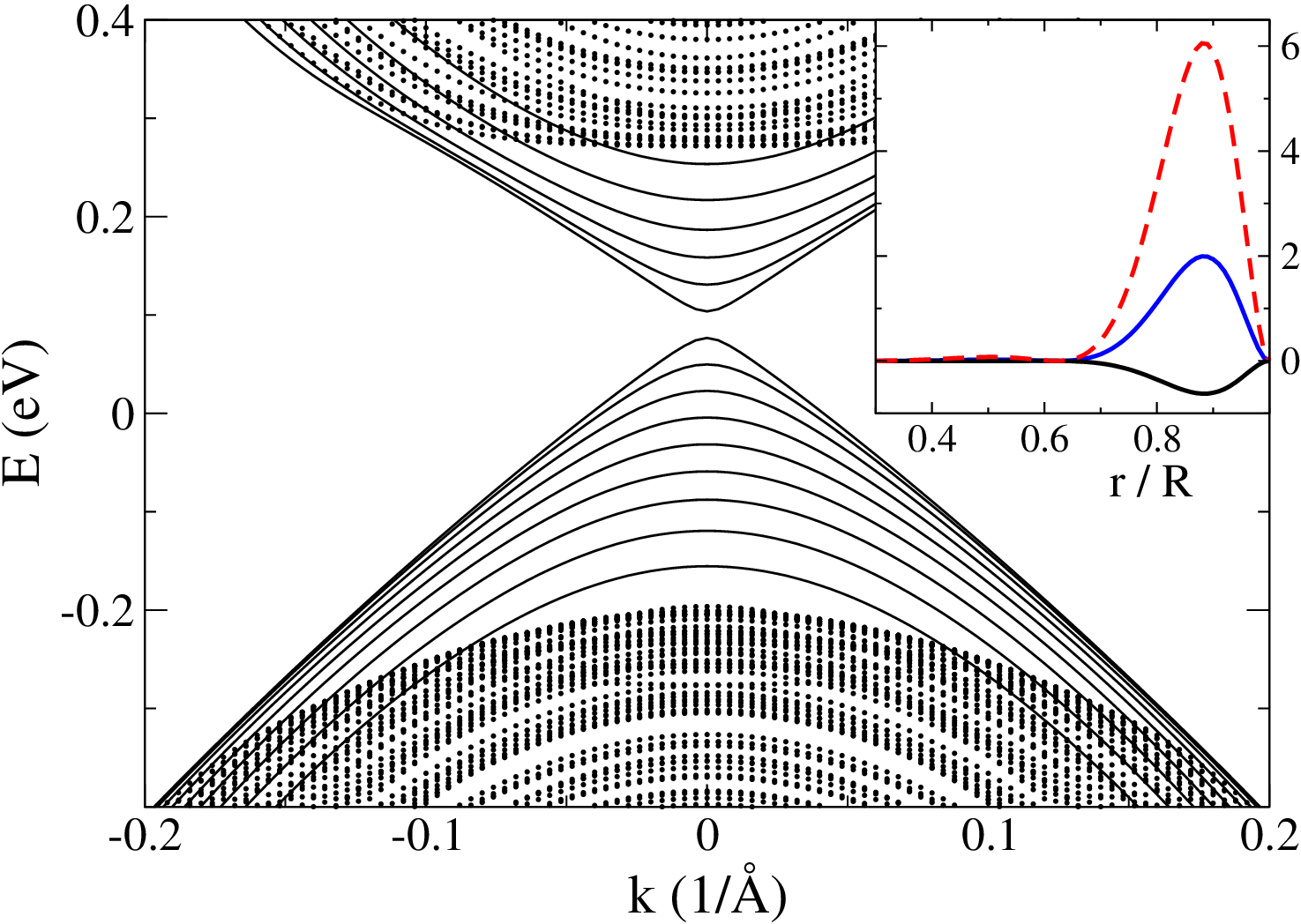}}
\caption{\label{f1} (Color online)  Band structure of a TI nanowire with 
$R=15$~nm obtained by numerical diagonalization  of Eq.~(\ref{hzah}).
Points refer to bulk states, lines to surface states.
Inset:  Density $\langle\rho\rangle$ (dashed red) and spin density 
[$\langle s_\phi\rangle$: blue, $\langle s_z\rangle$: 
black curve] vs radial coordinate for the right-moving
state $(k,j)=(0.02$~\AA$^{-1}, 1/2)$.}
\end{figure}

Let us start with the band structure of a clean noninteracting
cylindrical nanowire for $\Phi=0$. First, we 
employ the low-energy approach of Zhang \textit{et al.}
\cite{zhang} where, expanding up to order ${\bm k}^2$ in 
momentum around a suitable symmetry 
point, e.g., the $\Gamma$ point in Bi$_2$Se$_3$, 
the bulk TI Hamiltonian consistent with TRS plus inversion and rotation
symmetry has the form 
\begin{eqnarray}\label{h3d}
H_Z &=& \epsilon_0({\bm k}) \sigma_0\tau_0 + 
M({\bm k}) \sigma_0 \tau_z + A_1 k_z \sigma_z \tau_x \\ \nonumber &+& 
A_2 \tau_x (k_x \sigma_x + k_y\sigma_y).
\end{eqnarray}
The $z$ direction defines the anisotropy axis, 
 $\epsilon_0({\bm k}) = C+D_1k_z^2+ D_2 k_\perp^2$, 
$M({\bm k}) = M_0 + B_1 k_z^2+B_2 k_\perp^2$, 
and $k_\perp^2=  k_x^2+k_y^2$. We use Pauli matrices ${\bm \sigma}$
for spin and ${\bm \tau}$ for parity (orbital) space; 
$\sigma_0$ and $\tau_0$ denote the identity.
The TI phase is realized for $M_0 B_{1,2}<0$, and we take parameters
for Bi$_2$Se$_3$ as quoted in Ref.~\cite{zhang}.
For a nanowire along the $\hat e_z$ direction [we use cylindrical
 coordinates with unit vectors $\hat e_r=(\cos\phi,\sin\phi,0)$
and $\hat e_\phi=(-\sin\phi,\cos\phi,0)$, and put $\hbar=1$],
rotation symmetry in the $xy$ plane implies conservation of
total angular momentum $J_z=-i\partial_\phi + \sigma_z/2$,
with half-integer eigenvalues $j$.
For nanowire radius $R$, we require the wavefunction to 
vanish at the boundary $r=R$, which is automatically ensured    
by expanding in the orthonormal set of radial functions \cite{alfredo}
\[
u_{mn}(r<R) = \frac{\sqrt{2}}{R J_{m+1}\left(\gamma_{mn}\right)} \ 
J_{m}\left(\gamma_{mn} \frac{ r}{R}\right),
\]
where $\gamma_{mn}$ is the $n$th zero of the Bessel function $J_m$
with integer $m$.  For given $(k \equiv k_z,j)$, we express
$H_Z$  in the basis $|n,\sigma,\tau\rangle$,
where $\sigma=\pm$  ($\tau=\pm$) denotes the
eigenvalue of $\sigma_z$ ($\tau_z$) and 
the associated radial function is $u_{j-\sigma/2, n}(r)$.
Some algebra gives
\begin{eqnarray}\label{hzah}
H_Z |n\sigma\tau\rangle &=& ( \epsilon_0(k) + M(k) \tau)
|n\sigma\tau\rangle
+ A_1 k\sigma|n,\sigma,-\tau\rangle\\ \nonumber & +& \frac{2iA_2}{R}
\sum_{n'} \frac{\gamma_{j+\sigma/2,n'}\gamma_{j-\sigma/2,n} }{
\gamma^2_{j+\sigma/2,n'}-\gamma^2_{j-\sigma/2,n}} |n',-\sigma,-\tau \rangle
\end{eqnarray}
with the substitution $k_\perp\to \gamma_{j-\sigma/2,n}/R$ in 
$\epsilon_0({\bm k})$ and $M({\bm k})$.
Numerical diagonalization is then straightforward and 
yields topologically protected surface modes.
A typical band structure and spin (particle) density profiles are
shown in Fig.~\ref{f1}.  Evidently all surface modes have a finite
gap.
States with $(k,j)$ and $(-k,-j)$ form a Kramers degenerate pair,
and for given $k$, the $\pm j$ states are degenerate but have 
opposite $s_z$ spin polarization. 
We observe that the expectation values of
the spin density operators $s_{\phi,r,z}\equiv 
\frac12\hat e_{\phi,r,z}\cdot {\bm \sigma}$ 
only depend on the radial coordinate $r$.  
Since then $\langle s_r\rangle =-\partial_\phi\langle s_\phi\rangle =0$,
spin is always oriented tangential to the surface.
Moreover, we find that the spin direction 
always encloses the angle $\eta=\pi/2$ with the momentum
${\bm k}=k\hat e_z+(j/R) \hat e_\phi$. 
For large $|k|$,  a right (left) moving surface state
then has counter-clockwise (clockwise) 
spin polarization $\langle s_\phi\rangle >0$ ($\langle s_\phi\rangle<0$).

\begin{figure}
\scalebox{0.32}{\includegraphics{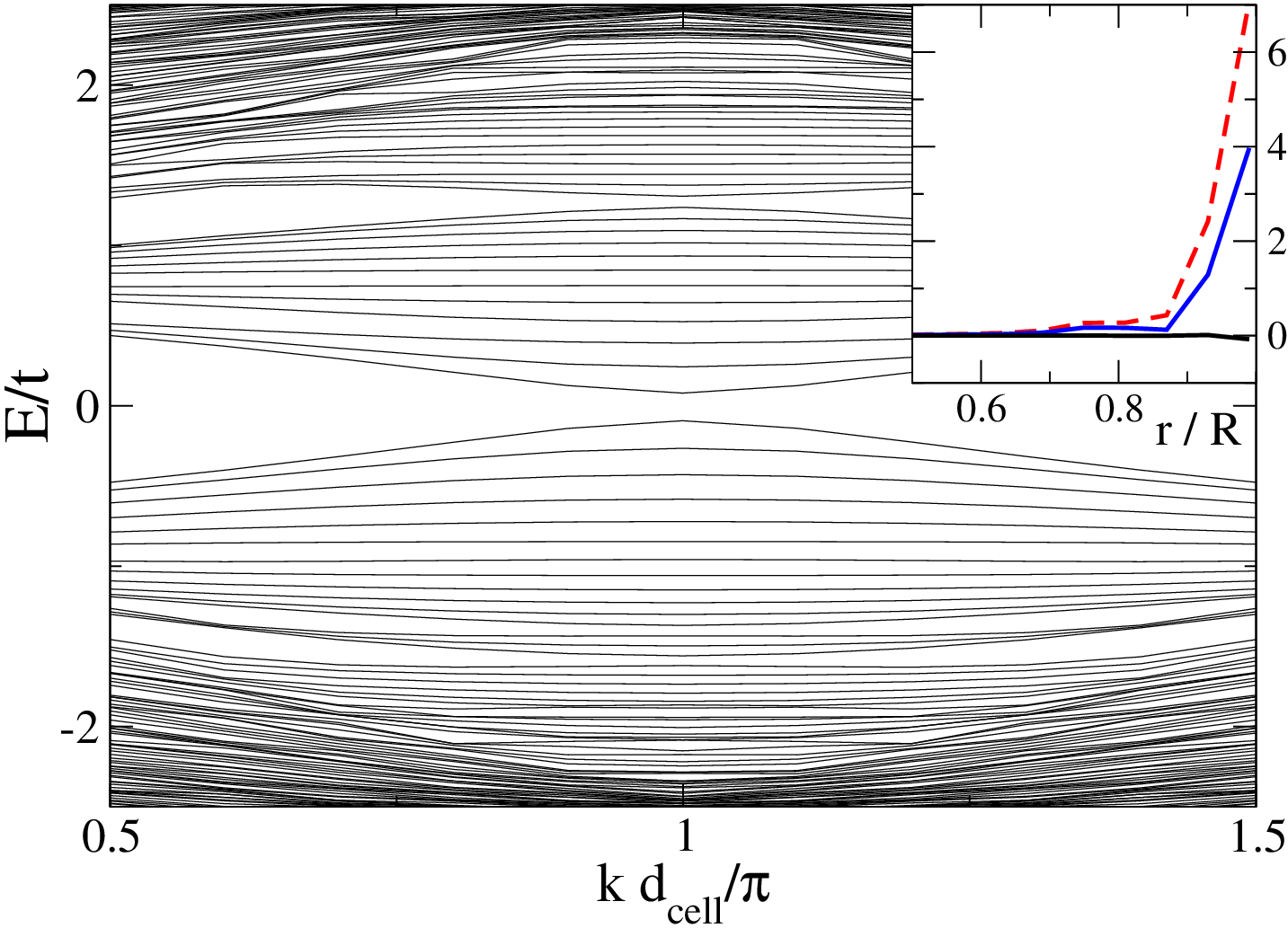}}
\caption{\label{f2} (Color online)  Same as Fig.~\ref{f1} but using
the tight-binding model on a diamond lattice, see Eq.~(\ref{tb}).  We take
$\hat e_z$ along the (111) axis,  $R=5a, \lambda_{\rm so}=
t,$ and $\delta t=0.28 t$. The size of the unit cell along this 
direction is $d_{\rm cell}= \sqrt{3} a$. 
Inset: as in Fig.~\ref{f1} but for $k d_{\rm cell}=2.9$.}
\end{figure}

\begin{figure}
\scalebox{0.32}{\includegraphics[angle=0]{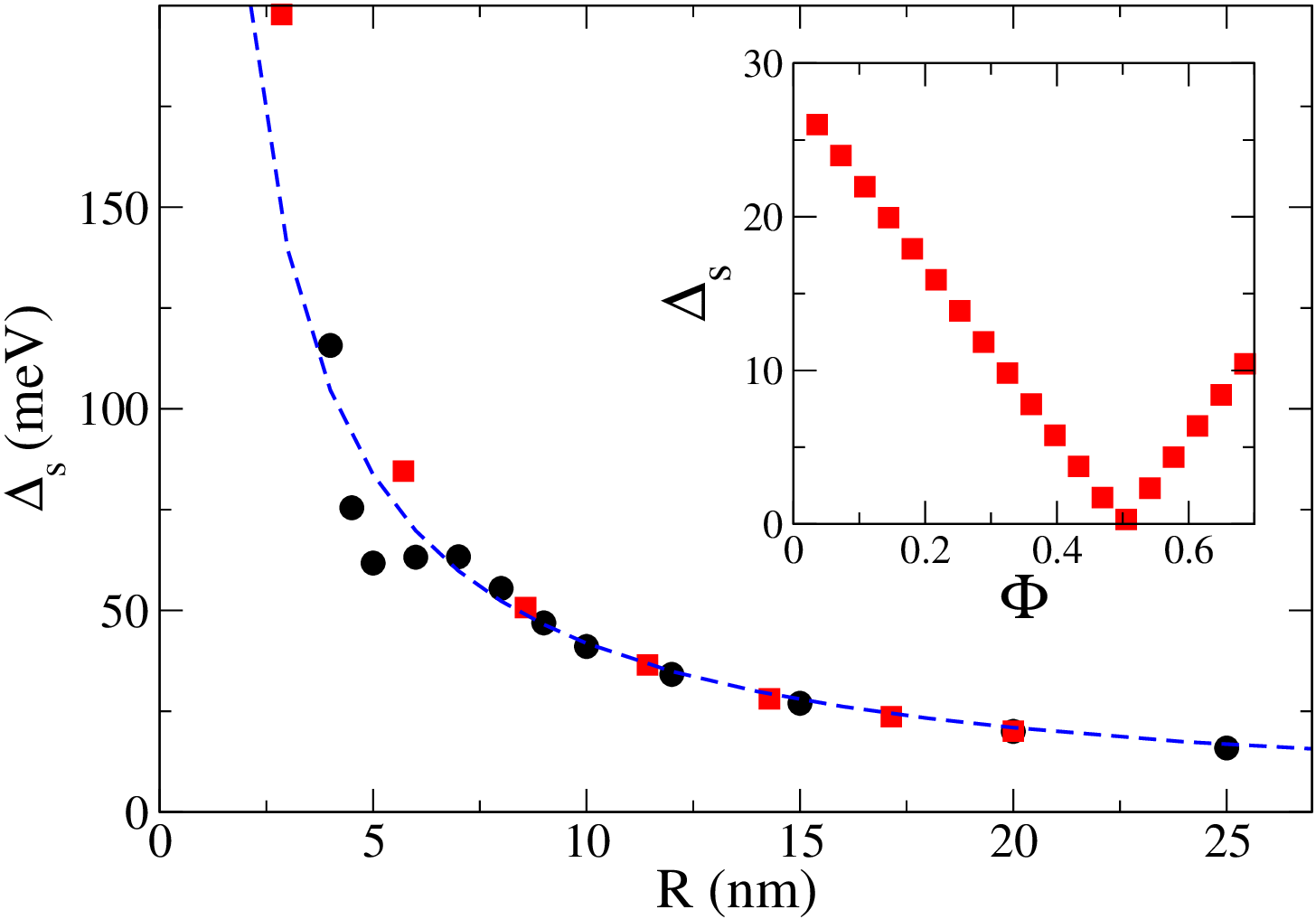}}
\caption{\label{f3} (Color online)  
Numerical results for the lowest surface state gap
$\Delta_s$ vs nanowire radius $R$, obtained  
from the low-energy approach (\ref{hzah}) [black circles] and from
the tight-binding model (\ref{tb}) [red squares].  The analytical 
prediction $v_2/R$, see  Eq.~(\ref{disp}), is given as blue dashed curve.
Inset: $\Phi$ dependence of $\Delta_s$ for $R=5a$ from the
tight-binding approach, where the flux is introduced via Peierls phases.}
\end{figure}

More microscopically, a TI nanowire can be described by a
tight-binding model for the electronic states in a diamond lattice with 
spin-orbit coupling $\lambda_{\rm so}$ \cite{fu},
\begin{equation}\label{tb}
H_{\rm tb} = \sum_{\langle i,j\rangle} t_{ij} c_i^\dagger c_j^{}
+ \frac{ 4i\lambda_{\rm so}}{a^2 }
\sum_{\langle\langle i,j\rangle\rangle}
c_i^\dagger \left( {\bm \sigma}\cdot  \left
[ {\bm d}_{ij}^1 \times {\bm d}_{ij}^2 \right] \right) c_j^{},
\end{equation}
where $a$ is the lattice spacing.  To reach the TI phase, 
a distortion $t_{ij}\to t_{ij}+\delta t$ in the 
nearest-neighbor hopping is introduced along the (111) 
direction. The $\lambda_{\rm so}$ term involves second neighbors and
depends on the two nearest-neighbor vectors ${\bm d}^{1,2}$ connecting them. 
After choosing an axis direction ($\hat e_z$), the wire is formed by 
all lattice sites located within radius $R$.  
A typical band structure for a nanowire with $R=5a$
 and $\hat e_z$ in the (111) direction
is shown in Fig.~\ref{f2}.  The surface states are again characterized by 
a finite gap, and spin or particle densities are qualitatively consistent with 
those shown in the inset of 
Fig.~\ref{f1}. The radius dependence of the lowest
surface state gap, $\Delta_s(R)$, obtained under both
approaches is compared in Fig.~\ref{f3}, where
 we set the parameter $-2t+\delta t$ in $H_{\rm tb}$, which is half
the gap in the (111) direction, 
 equal to $M_0$ in Eq.~\eqref{h3d}.  Agreement between both models
 at large $R$ is reached by adjusting $a$ to $2.8$~nm.
We see that even for very thin 
nanowires, the analytical prediction $\Delta_s=v_2/R$, see
Eq.~(\ref{disp}) below, agrees very well 
with the tight-binding result, while the 
low-energy model (\ref{hzah}) gives deviations when $R<5$~nm. 
It is worth mentioning that even though
a ${\bm k}\cdot {\bm p}$ expansion of $H_{\rm tb}$ 
around the $L = \frac{\pi}{a}(1,1,1)$ 
point does not match completely with Eq.~(\ref{h3d}),
the main features of the surface states are equivalent
in both descriptions. We have checked that similar 
results are obtained when $\hat e_z$ points along
 other crystallographic directions.

Both the dispersion relation and the spin texture of the surface states 
found under these two approaches are well reproduced by a model of
2D massless Dirac fermions wrapped onto the cylinder surface,
under the condition that the spin is tangential to the
surface and perpendicular to the momentum ($\eta=\pi/2$),  
cf.~Ref.~\cite{ran}.  To match the above numerical results, we also need to 
take into account anisotropy, since there are different Fermi velocities $v_{1,2}$ 
along the $\hat e_{z,\phi}$ directions. By supplementing
 Eq.~(\ref{h3d}) with boundary conditions describing a flat 2D surface 
in the $xz$ plane, we find $v_{1,2}=A_{1,2}\sqrt{1-(D_2/B_2)^2}.$
Taking the parameters of Ref.~\cite{zhang} for Bi$_2$Se$_3$, 
 $v_2/v_1\approx 2$.  
With $\eta=\pi/2$, the surface Hamiltonian takes the form \cite{ran}
\begin{equation}\label{surf}
H_{\rm surf} = e^{-i\sigma_z \phi/2} \left(
v_1 k \sigma_y - \frac{v_2}{R} \sigma_z (-i\partial_\phi+\Phi) \right)
e^{i\sigma_z\phi/2},
\end{equation}
where we added the dimensionless flux parameter $\Phi$.  
We note that $\Phi$ may include not only the orbital magnetic field, 
but also a Zeeman field or an exchange-coupled magnetization due to a nearby magnet (for fields along $\hat e_z$). 
The dispersion relation implied by Eq.~\eqref{surf} is
\begin{equation}\label{disp}
E_{k,j,\pm} = \pm \sqrt{ v^2_1 k^2+ \frac{v_2^2 ( j + \Phi )^2}{R^2} }  ,
\end{equation}
where the $\pm$ sign refers to conduction and valence band, respectively.
The corresponding eigenstate is
\begin{eqnarray}\nonumber
\psi_{k,j,\pm}(z,\phi) &\sim& e^{ikz+ij\phi} e^{-i\sigma_z\phi/2}
\left( \begin{array}{c} u_{k,j,\pm} \\ \pm i \sqrt{1-u_{k,j,\pm}^2}
\end{array} \right), \\   
u_{k,j,\pm} &=& \frac{v_1 k}{\sqrt{2E_{k,j,\pm}(E_{k,j,\pm}+(j+\Phi)v_2/R)}} .
\label{eigenstate}
\end{eqnarray}
For  integer $\Phi$, all bands are doubly degenerate and have 
a gap $\ge \Delta_s= v_2/R$. The mass term in the relativistic dispersion
(\ref{disp}) comes from a Berry phase $\pi$ due to 
spin-surface locking \cite{ran}.  While scattering between Kramers pairs,
$(k,j+\Phi) \leftrightarrow (-k,-(j+\Phi))$, is forbidden 
since the states (\ref{eigenstate}) have zero overlap,
 backscattering ($k\to -k$) for fixed $j$ is allowed, i.e.,
potential scattering (disorder) is relevant.
A non-integer flux $\Phi$ lifts the degeneracy, 
and for half-integer $\Phi$, 
the mass appearing in Eq.~(\ref{disp}) vanishes for the
special band $j=-\Phi$. This feature also appears
in the tight-binding calculation, see inset in Fig.~\ref{f3}.
Spin-conserving single-particle backscattering processes are
then forbidden, and weak disorder has no effect \cite{foot}. 
When the chemical potential $\mu$ is inside the bulk gap, we are thus
guaranteed to have an \textit{odd}\ number of modes.  

In the remainder, we focus on half-integer $\Phi$.
For simplicity, we consider $\mu<\Delta_s$ and sufficiently weak
interactions, where only the single mode
$j=-\Phi$ needs to be retained in a low-energy effective theory.
Moreover, we assume $\mu>0$ so that
Umklapp e-e scattering can also be neglected, cf.~Ref.~\cite{helical1}. 
Using the spinors (\ref{eigenstate}),
the surface electron operator $\Psi(z,\phi)$ is expanded in terms of 
slowly varying chiral 1D fermions $\psi_{r=\pm}(z)$, 
\begin{equation}\label{expand}
\Psi(z,\phi) = \frac{1}{\sqrt{4\pi}} \sum_{r=\pm} e^{irk_F z} \psi_r(z)
\left(\begin{array}{c} r \\ ie^{i\phi} \end{array}\right),
\end{equation}
with the Fermi momentum $k_F\equiv \mu/v_1$.
The standard bosonization approach \cite{gogolin} expresses 
$\psi_r(z)\simeq (2\pi \xi)^{-1/2} \exp[i\sqrt{\pi}(\varphi+r\theta)]$
in terms of conjugate phase fields $\varphi(z)$ and $\theta(z)$,  
where the surface layer width $\xi$ is the short distance cutoff 
for the 1D continuum description.  The noninteracting Hamiltonian is
$H_0=\frac{v_1}{2} \int dz [ (\partial_z\varphi)^2+(\partial_z\theta)^2]$.
The density operator, $\rho(z,\phi)=\Psi^\dagger \Psi$, is then equal to the
1D density,  $\partial_z\theta/\sqrt{\pi}$. Similarly,  
the spin density operators $s_\phi$ and $s_z$ are reduced to a 1D form,
\begin{eqnarray}\nonumber
\left(\begin{array}{c} s_\phi \\ s_z \end{array}\right)
&\equiv& \frac12 \Psi^\dagger(z,\phi)  
\left(\begin{array}{c} \sigma_y e^{i\phi\sigma_z} \\ \sigma_z
 \end{array}\right) \Psi(z,\phi) \\ \label{operator} & =& 
\left(\begin{array}{c} J(z) \\ 
-\frac{1}{\pi\xi}\cos[2k_F z+2\sqrt{\pi}\theta(z)] \end{array}\right).
 \end{eqnarray}
We observe that $s_\phi$ equals the 1D current density,
 $J(z)\equiv \partial_z\varphi/\sqrt{\pi}$, reflecting spin-momentum locking.  
There are no $2k_F$ oscillatory terms in $\rho$ nor in 
$s_\phi$.  On the other hand, no ``slow''
terms exist for $s_z$, and we always have $\langle s_z\rangle=0$. 

We now include e-e interactions, assuming that no metallic gates are nearby.
Similar to the nanotube case, apart from a hard-core part, 
the main contribution to the (surface-projected) 
potential can be modeled by \cite{cnt2}
\[
U ({\bm r}-{\bm r}') = 
\frac{e^2/\kappa}{\sqrt{(z-z')^2 + \xi^2+4R^2\sin^2[(\phi-\phi')/2]}},
\]
where $\kappa$ takes into account the dielectric constant of the
substrate and of the insulating interior part of the nanowire.  
Inserting the field operator (\ref{expand}) into the general 
second-quantized interaction Hamiltonian yields the 1D expression
\begin{equation}\label{ee}
H_{e-e} = \frac{1}{2\pi} \int dz dz' \ V(z-z')  \
\partial_z \theta(z) \partial_{z'}\theta(z')  
\end{equation}
with the effective 1D potential $V(z)=(2\pi)^{-1}\int_0^{2\pi} d\phi\
 U(z,\phi)$. The explicit form of $V$ is given in Ref.~\cite{cnt2} and
has the Fourier transform 
$\tilde V(q) \simeq (2e^2/\kappa)[0.51- \ln|qR|]$
for $|q|R \ll 1$.
Hard-core interaction terms give an additional 
contribution $b\int dz [(\partial_z\theta)^2-(\partial_z\phi)^2 ]$ to
the Hamiltonian, where $b$ depends on microscopic details.   
Since $b$ stays marginal under renormalization group
transformations, the logarithmic singularity in $\tilde V(q)$, caused by
the long-ranged Coulomb tail, is expected to dominate in practice.
Approximating $q\approx 2\pi/L$ for nanowire length $L$, 
we obtain the single-mode helical LL \cite{helical1},
$H_{\rm hLL} = \frac{v}{2}\int dz 
\left[ K (\partial_z\phi)^2+ K^{-1}(\partial_z\theta)^2\right]$,
where $v=v_1/K$ and
\begin{equation}\label{kk}
K = \frac{1}{\sqrt{1+\frac{2e^2}{\pi \kappa v_1} ( \ln[L/(2\pi R)] + 0.51)}},
\end{equation}
with $K=1$ without interactions.  It is straightforward
to generalize these expressions to $b\ne 0$ \cite{gogolin,helical3}.  
For $L/R\approx 1000$, we find 
the strongly correlated value $K\approx 0.4$ to $0.5$ from Eq.~(\ref{kk}).
We note that $K\approx 0.53$ to $0.9$ \cite{helical3}
for the QSH edge in HgTe/CdTe wells. 

The helical LL state in a TI nanowire can be identified through several
experimentally observable signatures.
First, we note that the equal-time spatial correlations of $\rho$ and 
$s_\phi=J$ decay as $|z|^{-2}$.  While $\langle s_z\rangle=0$ remains
valid for arbitrary $K$, 
$s_z$ correlations show a slow algebraic power-law decay,
 $\langle s_z(z) s_z(0) \rangle \propto \cos(2k_F z) |z|^{-2K}$. 
For $K<1$,  we therefore find an ordering tendency towards 
spin density wave (SDW) formation, where spins are oriented along 
the nanowire axis $\hat e_z$.  Within the standard classification of 1D 
systems \cite{gogolin}, the helical LL in a TI nanowire is thus in a
SDW phase.  As a consequence, the Ruderman-Kittel 
interaction among magnetic impurities mediated by such a nanowire 
is extremely anisotropic (see also Ref.~\cite{rkky}) and decays
only with a slow power law. At the same time, the absence of $2k_F$ terms
in the density operator implies that no charge density wave (CDW) correlations
develop at all. Furthermore, the superconducting order parameter
describing  singlet Cooper pairing is ${\cal O}(z,\phi)\sim 
e^{i\phi} \psi_+(z)  \psi_-(z)$,
which implies a fast power-law decay $\propto |z|^{-2/K}$.
The angular $e^{i\phi}$ dependence comes from the spin structure in
Eq.~(\ref{expand}) and causes an additional
 strong suppression of the proximity effect
when bulk superconductors are in contact to the nanowire.
For normal-state metallic electrodes, in a two-terminal geometry,
the conductance is $G=e^2/h$ (independent of $K$) when 
the contacts are ideal.  However,
non-ideal contacts cause a typical temperature-dependent
decrease of $G(T)$ at low temperatures due to the well-known power-law 
suppression of the tunneling density of states \cite{helical3}.
Moreover, in contrast to a spin-polarized LL (which also has
$G=e^2/h$ for ideal contacts), spin plays an essential role here.
This could be easily seen in the presence of  magnetic impurities. 
In particular, the Kondo effect can take place, where 
theoretical predictions for $G(T)$ exist \cite{kondo} and directly apply.  
To conclude, we are confident that the helical LL will soon allow
for its clear experimental identification in topological insulator nanowires.

This work was supported by the SFB TR 12 of the DFG,
the ESF network INSTANS, and by the Spanish
MICINN under contract FIS2008-04209.


\begin{thebibliography}{99}

\bibitem{hasan}
For reviews see:
 X.-L. Qi and S.-C. Zhang, Phys. Today {\bf 63}, 33 (2010);
 M.Z. Hasan and C.L. Kane, arXiv:1002.3895.

\bibitem{surface}
Y. Xia \textit{et al.}, Nat. Phys. {\bf 5}, 398 (2009).

\bibitem{butch}
N.P. Butch \textit{et al.}, Phys. Rev. B {\bf 81}, 241301(R) (2010).

\bibitem{chch}
J.G. Checkelsky, Y.S. Hor, R.J. Cava, and N.P. Ong, arXiv:1003.3883.

\bibitem{exp1}
H. Peng \textit{et al.},  Nat. Mat. {\bf 9}, 225 (2009).

\bibitem{exp2}
D. Kong \textit{et al.}, Nano Lett. {\bf 10}, 329 (2010). 

\bibitem{ashvin}
Y. Zhang and A. Vishwanath, arXiv:1005.3542.

\bibitem{piet}
J.H. Bardarson, P.W. Brouwer, and J.E. Moore, arXiv:1005.3762.

\bibitem{zhang}
H. Zhang,  C.-X. Liu, X.-L. Qi, X. Dai, Z. Fang, 
and S.-C. Zhang, Nat. Phys. {\bf 5}, 438 (2009).

\bibitem{fu}
L. Fu, C.L. Kane, and E.J. Mele, Phys. Rev. Lett. {\bf 98}, 106803 (2007).

\bibitem{ran}
Y. Zhang, Y. Ran, and A. Vishwanath, Phys. Rev. B {\bf 79}, 245331 (2009).

\bibitem{cnt1}
R. Egger and A.O. Gogolin, Phys. Rev. Lett. {\bf 79}, 5082 (1997);
C.L. Kane, L. Balents, and M.P.A. Fisher, Phys. Rev. Lett. {\bf 79}, 
5086 (1997).

\bibitem{cnt2}
 R. Egger and A.O. Gogolin, Eur. Phys. J. B {\bf 3}, 281 (1998).

\bibitem{gogolin}
A.O. Gogolin, A.A. Nersesyan, and A.M. Tsvelik, 
\textit{Bosonization and strongly correlated systems}
(Cambridge University Press, 1998).

\bibitem{mora}
T. Giamarchi and H.J. Schulz, Phys. Rev. B {\bf 37}, 325 (1988);
C. Mora, R. Egger, and A. Altland, Phys. Rev. B {\bf 75}, 035310 (2007).

\bibitem{helical1}
C. Wu, B.A. Bernevig, and S.-C. Zhang, Phys. Rev. Lett. {\bf 96}, 106401 (2006).

\bibitem{helical2}
C. Xu and J.E. Moore, Phys. Rev. B {\bf 73}, 045322 (2006).


\bibitem{konig}
M. K\"onig \textit{et al.}, J. Phys. Soc. Jpn. {\bf 77}, 031007 (2008).

\bibitem{helical3}
C.Y. Hou, E.A. Kim, and C. Chamon, Phys. Rev. Lett. {\bf 102}, 076602 (2009);
A. Str\"om and H. Johannesson, Phys. Rev. Lett. {\bf 102}, 096806 (2009);
J.C.Y. Teo and C.L. Kane, Phys. Rev. B {\bf 79}, 235321 (2009). 

\bibitem{alfredo}
N. Agrait, A. Levy Yeyati, and J.M. van Ruitenbeek, Phys. Rep.
{\bf 377}, 81 (2003). 

\bibitem{foot}
Note that two-particle backscattering effects combined with very strong 
interactions can destabilize the helical liquid \cite{helical1,helical2}. 

\bibitem{rkky}
J. Gao, W. Chen, X.C. Xie, and F.-C. Zhang, Phys. Rev. B {\bf 80}, 241302(R)
(2009).

\bibitem{kondo}
J. Maciejko, C. Liu, Y. Oreg, X.-L. Qi, C. Wu, and S.-C. Zhang,
Phys. Rev. Lett. {\bf 102}, 256803 (2009).

\end{thebibliography}
\end{document}